\documentclass[a4paper,fleqn,usenatbib]{mnras}

\usepackage{newtxtext,newtxmath}

\usepackage[T1]{fontenc}
\usepackage{ae,aecompl}

\usepackage{graphicx}	
\usepackage{amsmath}	
\usepackage{amssymb}	

\newcommand{\um }{$\mu $m}
\newcommand{\ergsec}{erg~s$^{-1}$~cm$^{-2}$~arcsec$^{-2}$}
\newcommand{\SB}{erg~s$^{-1}$~cm$^{-2}$~arcsec$^{-2}$}
\newcommand{\EM}{cm$^{-6}$~pc}
\newcommand{\cmq}{cm{$^{-3}$}}

\newcommand{\kms}{km~s{$^{-1}$}}

\newcommand{\Msol}{M{$_{\odot}$}}

\newcommand{\Lsol}{L{$_{\odot}$}}

\newcommand{\Hii}{H{\sc ii}}
\newcommand{\Nii}{N{\sc ii}}
\newcommand{\Sii}{S{\sc ii}}
\newcommand{\Oi}{O{\sc i}}
\newcommand{\Oii}{O{\sc ii}}
\newcommand{\Oiii}{O{\sc iii}}
\newcommand{\Ha}{H$\alpha$}

\title[Red Square jet]{A Highly Collimated Jet from the Red Square Nebula, MWC~922}

\author[J. Bally et al.]{
John Bally,$^{1}$\thanks{E-mail: john.bally@colorado.edu}
Zen H. Chia,$^{1}$
\\
$^{1}$Center for Astrophysics and Space Astronomy, Department of Astrophysical and Planetary Sciences, \\
   University of Colorado, Boulder, CO 80389, USA \\
}

\date{Accepted XXX. Received YYY; in original form 25 October 2018}

\pubyear{2015}

\begin{document}
\label{firstpage}
\pagerange{\pageref{firstpage}--\pageref{lastpage}}
\maketitle

\begin{abstract}
Deep, narrow-band \Ha\  and 6584 \AA\ [\Nii ] CCD images 
of the peculiar, infrared excess B[e] star  MWC~922 reveal a collimated, 
bipolar  jet  orthogonal to the previously detected extended nebula. 
The jet consists of a pair of $\sim$0.15 pc segments on either side of  
MWC~922 separated by gaps.  The most distant jet segments disappear 
$\sim$0.6 pc from the  star.  The northwest beam points to 
a faint emission-line feature 1.65 pc from MWC~922  that may be a terminal 
bow shock where the jet rams the ambient medium.   The narrow
opening-angle of the jet combined with an estimated internal sound speed 
of $\sim$10 \kms\ implies a jet speed $\sim$500 \kms .    The previously 
detected nebula extends up to 0.6~pc to the southwest of MWC~922 
at right angles  to the jet and appears to be an extension of the compact, 
edge-on disk surrounding the star.  It points toward the  HII region 
Messier 16 located $\sim$1\degr\  ($\sim$30 pc in projection) to the 
southwest.   This nebula and jet appear to be externally ionized 
by the ambient Lyman continuum radiation field and have electron densities  
of n$_e \sim$ 50 to 100 cm$^{-3}$.  The southwest nebula and jet have 
similar surface brightness in \Ha\ and [\Nii ].   Faint 70 $\mu$m  
emission traces the southwest ejecta that  likely originates
from $\sim$50 K dust  embedded in the photo-ionized plasma 
which may shadow the dimmer ejecta  northeast of MWC~922.  
MWC~922 may be a massive member of the  Serpens OB1 or 
OB2 associations  surrounding Messier 16 and Sh2-54.   
\end{abstract}




\section{Introduction}

\begin{figure*}
	\includegraphics[width=\columnwidth]{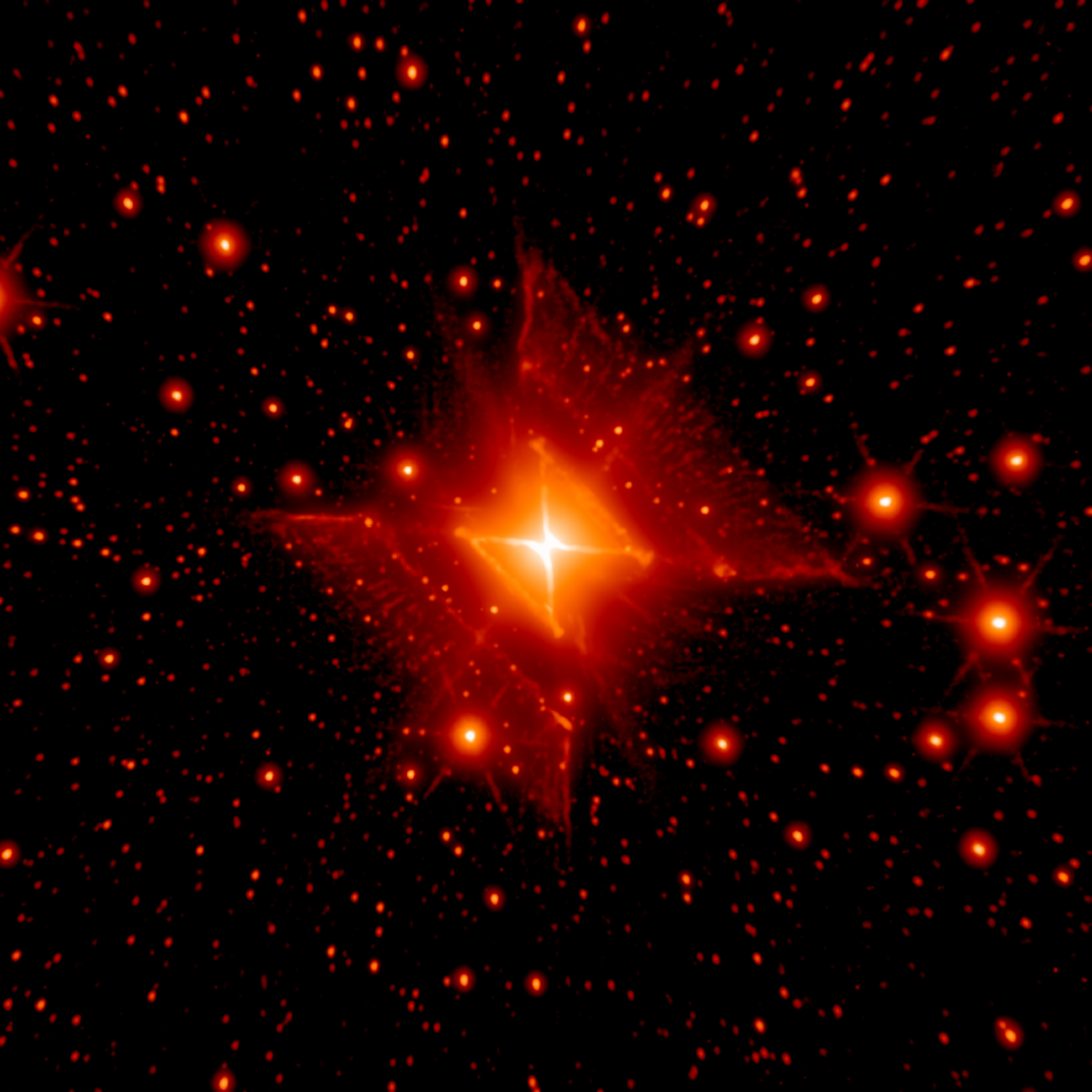}
    \caption{The Red Square Nebula showing an H-band adaptive optics image 
    combining data from the Palomar Observatory 200-inch telescope which
    was published in  \citet{Tuthill_Lloyd2007} with adaptive optics data from the
    Keck Observatory 10-meter.  North is up and East is to the left. 
    This image was featured on APOD on 31 January 2016.  See
    https://apod.nasa.gov/apod/ap160131.html.  Image Credit: 
    Peter Tuthill, Sydney University Physics Dept., and the Palomar and W.M. Keck 
    observatories.}
    \label{fig1}
\end{figure*}

MWC~922 (IRAS 18184-1302), also known as the {\it Red Square Nebula}  (RSN) 
based on its  remarkable X-shaped near-infrared morphology \citep{Tuthill_Lloyd2007},  
is a dust/gas-enshrouded emission line star with strong infrared excess at
mid-IR wavelengths (see Figure \ref{fig1}).    
The spectrum of MWC~922 is dominated by strong emission 
in hydrogen recombination lines, a forest of fainter [Fe I] and [Fe II] lines, molecular 
emission and absorption features, and some diffuse interstellar bands \citep{Wehres2017}.

\citet{Lamers1998} classified  B[e] stars into various subgroups, but catalogued  
MWC~922 as an  {\it unclassified} 'uncl~B[e]'  star owing to its peculiar 
atomic, ionic, and molecular spectral features.   Some B[e] stars are 
young and closely associated with their parent molecular clouds, and others 
appear to be main-sequence multiples experiencing mass transfer onto a companion, 
while the rest appear to be  post-main-sequence objects.  \citet{Miroshnichenko2007}
clarified the nature of the `uncl~B[e]' stars,  and renamed members of this class as 
`FS CMa' stars.   FS CMa stars are binary systems that are currently undergoing, 
or have recently  undergone a phase of rapid mass exchange, strong mass loss, 
and dust formation.   However, the evolutionary stage of MWC~922 remains unclear.

\citet{Rodriguez2012} detected a bright, compact 3.6 cm radio source having dimensions
of 0.18\arcsec\ $\times$ 0.20\arcsec\ in the center of the RSN. 
\citet{Sanchez_Contreras2017}  detected radio recombination lines H30$\alpha$, 
H31$\alpha$ at $\lambda$ = 3 mm and H39$\alpha$, H41$\alpha$ 
at $\lambda$ = 1 mm along with several H radio-recombination lines in the 
$\beta$ and and $\gamma$ series.   From the
velocity centroids,  \citet{Sanchez_Contreras2017}  measured the LSR radial velocity
of MWC922 to be V$_{LSR}$ = +32.5$\pm$0.4 \kms .    
The radial velocity combined
with the Galactic rotation curve implies D = 1.7 to 3 kpc if located at the near distance
\citep{Sanchez_Contreras2017}.  

The bright, inner nebulosity surrounding MWC~922, which is about 10\arcsec\  
in diameter, is similar to the well-studied 
proto-planetary nebula (pPN) known as the {\it Red Rectangle} \citep{Bujarrabal2016}.
However, the symmetry of the RSN indicates that its disk and the biconical outflow
cavity oriented orthogonal to the disk are seen nearly edge-on.  Indeed, 
adaptive optics images in the near-IR (Figure \ref{fig1}) show a nearly 
edge-on equatorial dust lane  with a major axis at position angle PA = 46\degr .

\citet{Marston2008} detected an extended and elongated \Ha\ and [\Nii ] nebula 
associated with the RSN oriented northeast-southwest, with a major-axis 
dimension of at least 110\arcsec\ and a minor-axis dimension of about 
10\arcsec\  to 20\arcsec .     The brightest portion of this nebula is located 
southwest of  MWC~922 and points toward Messier  16 located about 1\degr\
($\sim$ 30 pc in projection) to the southwest at PA $\sim $ 217\degr .

If MWC~922 is related to the Messier 16 
star forming complex it may be a member of the Serpens OB1 association.  
If so, it is  likely to have a distance  between$\sim$~1.7  to 2.0 kpc.   
\citet{Guarcello2007}  investigated the stellar population 
associated with Messier 16 for which a distance of 1.75 kpc was found. 
Thus, a  value of 1.7 pc is adopted for the distance of MWC~922
in this study.  

The radial velocity of molecular gas in the vicinity of Messier 16 ranges 
from V$_{LSR}$ = 17 to 29 \kms , about $\sim$ 10 \kms\ lower than
the gas associated with MWC~922 \citep{Nishimura2017,Pound1998}.
A projected separation of 30 pc between MWC~922 and the core of Messier 16
and a radial velocity difference of 10 \kms\ either implies that the two 
objects are unrelated, or that MWC~922 is a low-velocity runaway star from
Serpens OB1. 

The Serpens OB2 association associated with the \Hii\ region Sh2-54 and 
NGC~6604 located about 1.5\degr\ to the northwest of MWC~922 is 
also at a similar distance \citep{Reipurth2008,Davidge1988}.  
The combined light from massive stars in Serpens OB1 and OB2 may be 
responsible for the extended diffuse \Ha\ emission  in the region and for 
external ionization of the nebulosity and jets associated with 
MWC~922.

The spectral energy distribution (SED) of MWC~922 \citep{Sanchez_Contreras2017} 
implies a luminosity of L $\sim$ 1.8$\times$10$^4$ \Lsol\ at D = 1.7 kpc after 
correction for foreground extinction, estimated to be between A$_V$ = 1.7 to 2.7 
magnitudes.  

\section{Observations}

\begin{table}
	\centering
	\caption{Observation Summary.}
	\label{table1}
	\begin{tabular}{lll} 
		\hline
          Date & Filter & Exposure time \\
		\hline
19 Aug. 2012	& DIS, R=300		& 3 $\times$ 600;  2 $\times$ 600 \\ 
22 Mar. 2017 	& DIS, R=1200  	& 2 $\times$ 600  \\ 
04 July 2016   	& UNM657$/$3  	& 3 $\times$ 300 \\ 
                       	& UNM659$/$3  	& 3 $\times$ 300;  3 $\times$ 900 \\
09 July 2016  	& UNM659$/$3  	& 4 $\times$ 600 \\ 
                       	& CU673$/$10     	& 3 $\times$ 300 \\
4 Sept. 2016   	& UNM659$/$3 	& 4 $\times$ 600 \\ 
5 Sept. 2016 	& UNM659$/$3 	& 5 $\times$ 900 \\
                		& UNM657$/$3 	& 9 $\times$ 300 \\  
13 May 2018 	& UNM657$/$3 	& 3 $\times$ 300s \\ 
                		& UNM659$/$3 	& 6 $\times$ 300s \\
                  \hline
	\end{tabular}
\end{table}

Several  R$\approx$300 spectra
covering 2500\AA -wide  spectral windows around H$\alpha$ and H$\beta$  were acquired with the Double Imaging Spectrograph (DIS) on the 3.5 meter telescope at the Apache Point Observatory (APO) located near Sunspot, New Mexico, USA on  19 August 2012.  A grating with a dispersion of 2.31\AA\ per pixel and a spatial scale of 0.4\arcsec\ per pixel was used. For three 600 second duration exposures, the 5\arcmin\  by 1.5\arcsec\ wide long-slit was centered on MWC922 and oriented at PA = 45\degr\ along the major axis of the elongated nebula surrounding the star.  A pair of 600 second exposures were also obtained at an offset position 5\arcsec\ northwest at the same position angle.  
These spectra show that the nebula extends over 1\arcmin\ southwest from MWC922 and is nearly equally bright in \Ha\ and the 6584\AA\ [\Nii ] emission line.     Consequently, we used the  APO 3.5 meter f/10 telescope to acquire narrow-band images that isolated the \Ha ,[\Nii ],  and the red [\Sii ] doublet.    An additional  pair of  R$\approx$3,000  spectra,  each covering 1160\AA\ spectral windows around H$\alpha$ and 1240 \AA\ around H$\beta$,  were obtained with DIS using  exposure times of 600 seconds on 22 March 2017.   The scale of these spectra is 0.58\AA\ per pixel in the red and 0.74\AA\ in the blue.

Narrow-band images were obtained between 4 July 2016 and 13 May 2018 using the ARCTIC CCD camera on the APO 3.5 meter telescope.   ARCTIC uses a backside-illuminated 4096$\times$4096 pixel chip with 15 $\mu$m pixel pitch,   an image scale of 0.114\arcsec\ per pixel in unbinned mode, and a 7.8\arcmin $\times$7.8\arcmin\ view of view.  To better match to the native seeing at APO, the observations reported here used 3$\times$3 binning to provide a scale of 0.344\arcsec\ per pixel, implying that there are $N_{pix}$ = 8.427 pixels per square-arc-second.  Table~1 summarizes the observation dates and exposure times.   Images were obtained using narrow-band filters having bandpasses of 30\AA\  centered on the 6563\AA\ H$\alpha$  and 6584\AA\ [NII] emission lines (UNM657/30, UNM659/30)  and a 100\AA\ bandpass filter centered on the red  $\lambda \lambda$6717/6731\AA\  [\Sii ] doublet (CU SII) with exposure times ranging from 300 to 900 seconds per frame.  The images were processed in the standard fashion using bias, dark, and twilight flat frames.     Between 2016 and 2018, a pattern of elevated dark current became apparent in the images.  While the 2016 data could be processed by subtraction of a median-filtered stack of bias frames, the 2018 image required subtraction of dark frames to remove these artifacts.  The dark-subtracted images were flat-fielded using twilight flats.    The processed, individual frames were de-distorted using 2MASS coordinates of stars in the field.  Final images in each filter were combined using a median combination of the registered frames.  

Approximate photometric calibration is based on the SDSS r-filter magnitudes of non-variable (at a few percent level), unsaturated,  in-field stars identified in the Pan-STARRS catalog.   For both the UNM657/30 and UNM659/30 filters, the zeropoint (defined as the magnitude of a star that produces 1 ADU per second per pixel if all the starlight were concentrated into one pixel) is ZPT = 20.93$\pm$0.1.   For the 100\AA\ bandpass [\Sii ] filter, ZPT=22.23$\pm$0.1.  These values were used to convert counts in the images to flux-densities in a given photometry-aperture (F$_{\nu}$ in Jy and erg~s$^{-1}$~cm$^{-2}$~Hz$^{-1}$)  using the flux-density of Vega (approximately 3174 Jy at the \Ha\ wavelength of 6563\AA\  and 3085 Jy at the mean wavelength of the red [\Sii ] doublet,  6724\AA  ).   Background-subtracted emission line surface brightness values (SB: erg~s$^{-1}$~cm$^{-2}$~arcsec$^{-2}$) are obtained by multiplication of the flux-density by the bandpasses of the filters ($\sim$ 30\AA\ for \Ha\  and [\Nii ] and $\sim$100\AA\ for [\Sii ]) and division by the areas (in square arc-seconds) of the photometry apertures.   Fluxes and surface brightness measurements were converted into estimates of the emission measures (EMs), electron densities, mass loss rates, momentum injection rates, and mechanical luminosities as discussed below.    

The images shown in Figures \ref{fig2} to \ref{fig5} have been converted to surface brightness units (\SB ) and adjusted to remove added flux from scattered light, light pollution,  and airglow.     The Southern H-alpha Sky Survey Atlas \citep[SHASSA; ][]{Gaustad2001}  was used to estimate the amount of added flux in our images.  After conversion, the data values in each pixel give the observed  surface brightness in \SB .
  
The conversion from data values to surface-brightness  units is achieved by multiplication by the following conversion constants:  $2.814 \times 10^{-16} N_{pix} / \tau _{exp}$ for \Ha\ and [\Nii ] and $8.499 \times 10^{-17} N_{pix} / \tau _{exp}$ for [\Sii ] where $ \tau _{exp}$ is the effective exposure time in seconds (300 seconds for \Ha , 900 seconds for [\Nii ], and 600 seconds for [\Sii ] in the median combined final images).

The added foreground flux is taken to be the difference between the continuum-subtracted SHASSA \Ha\ surface-brightness  and the diffuse \Ha\ surface-brightness  in our ARCTIC \Ha\ image.    This surface-brightness  value is subtracted from the final \Ha\ image.    The ARCTIC [\Nii ] and [\Sii ] images are corrected for the added foreground by assuming that the background emission in these emission lines is 0.3 and 0.12 times the \Ha\ surface-brightness as was determined from our DIS spectra. 

The surface brightness scales in  Figures \ref{fig2} to \ref{fig5} assume $A_V$ = 0 magnitudes of interstellar extinction.  However, for the analysis below, it is assumed that the extinction to Red Square Nebula is  $A_V$ = 2 magnitudes.  It is evident from the \Ha\ image that the extinction across the field is highly structured and variable.   Thus, the physical parameters are uncertain by at least a factor of two.
 
\section{Results}

The area-integrated flux ratio $I_{[\Nii  ]} / I_{H \alpha } $ is  0.8$\pm$0.1 (in the DIS spectra) in a region extending southwest of MWC~922 from 20\arcsec\ (0.16 pc) to 60\arcsec\  (0.5 pc) from the star.    At positions away from the Red Square Nebula (RSN), $I_{[\Nii  ]} / I_{H \alpha} $ is about 0.3$\pm$0.1 everywhere.   Thus, the [\Nii ] emission is enhanced relative to the \Ha\ light measured at locations where the slit crossed the RSN.    As discussed below, unlike in [\Nii ], there is no enhancement in the [\Sii ] emission from the RSN.  The $\lambda \lambda $6717/6731 line ratio is around 1.3$\pm$0.1 in the same DIS aperture in which the [\Nii ]/\Ha\ ratio is enhanced.   This is close to the low-density limit for the [\Sii ] line ratio and implies a density $n_e <$240 \cmq .    The mean [\Sii ] doublet ratio from the background nebula is also near the low-density limit, $\sim$1.3$\pm$0.1, implying a mean density  $<$240 \cmq .

\begin{figure*}
	\includegraphics[width=5in]{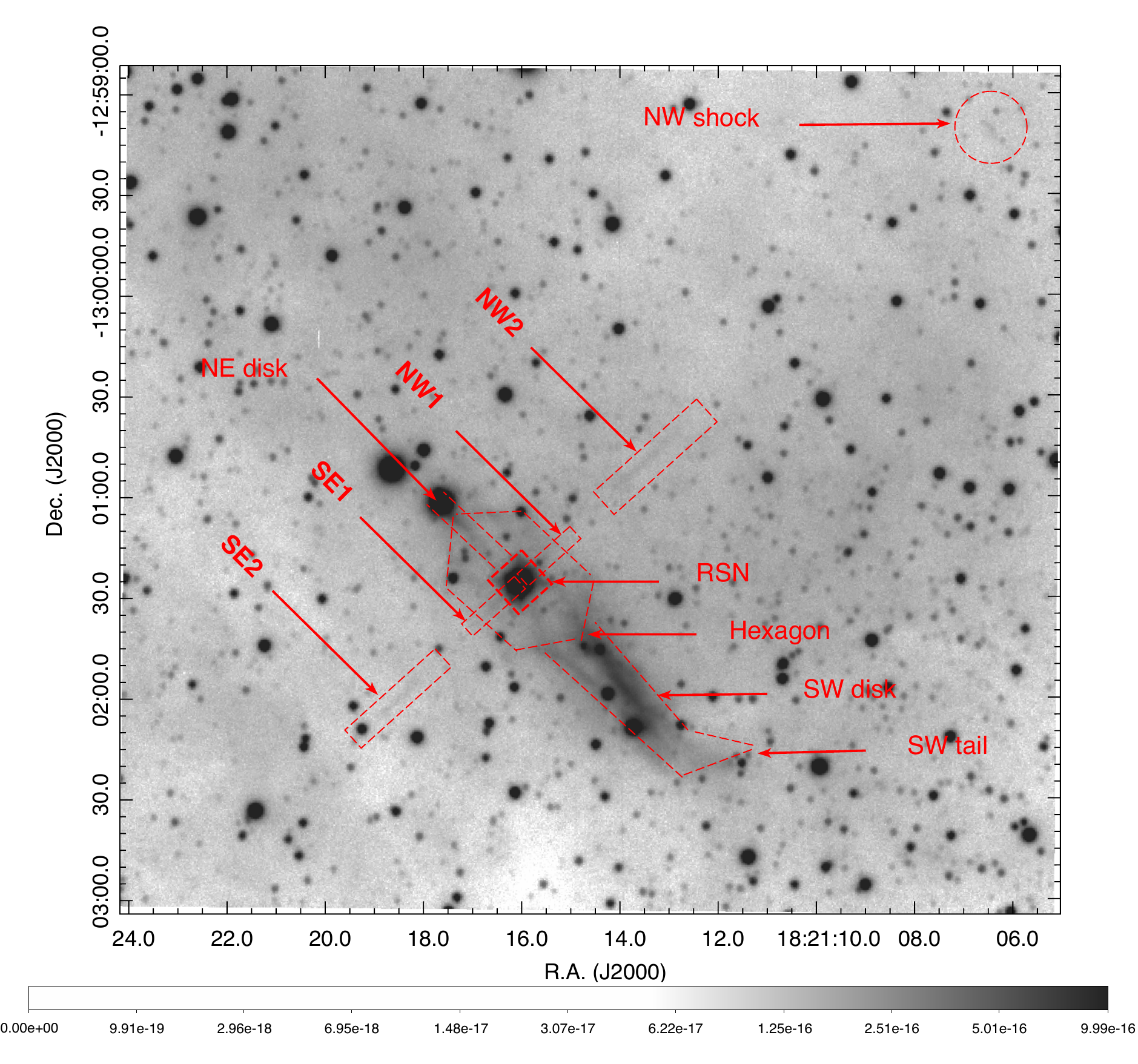}
    \caption{The Red Square Nebula showing the [\Nii ]  emission using a logarithmic display.  The displayed surface brightness ranges from $0$ to $\rm 1 \times 10^{-15}$ (\SB ).
    Labels indicate the various features discussed in the text and listed in Table~2.}
    \label{fig2}
\end{figure*}

\begin{figure*}
	\includegraphics[width=5in]{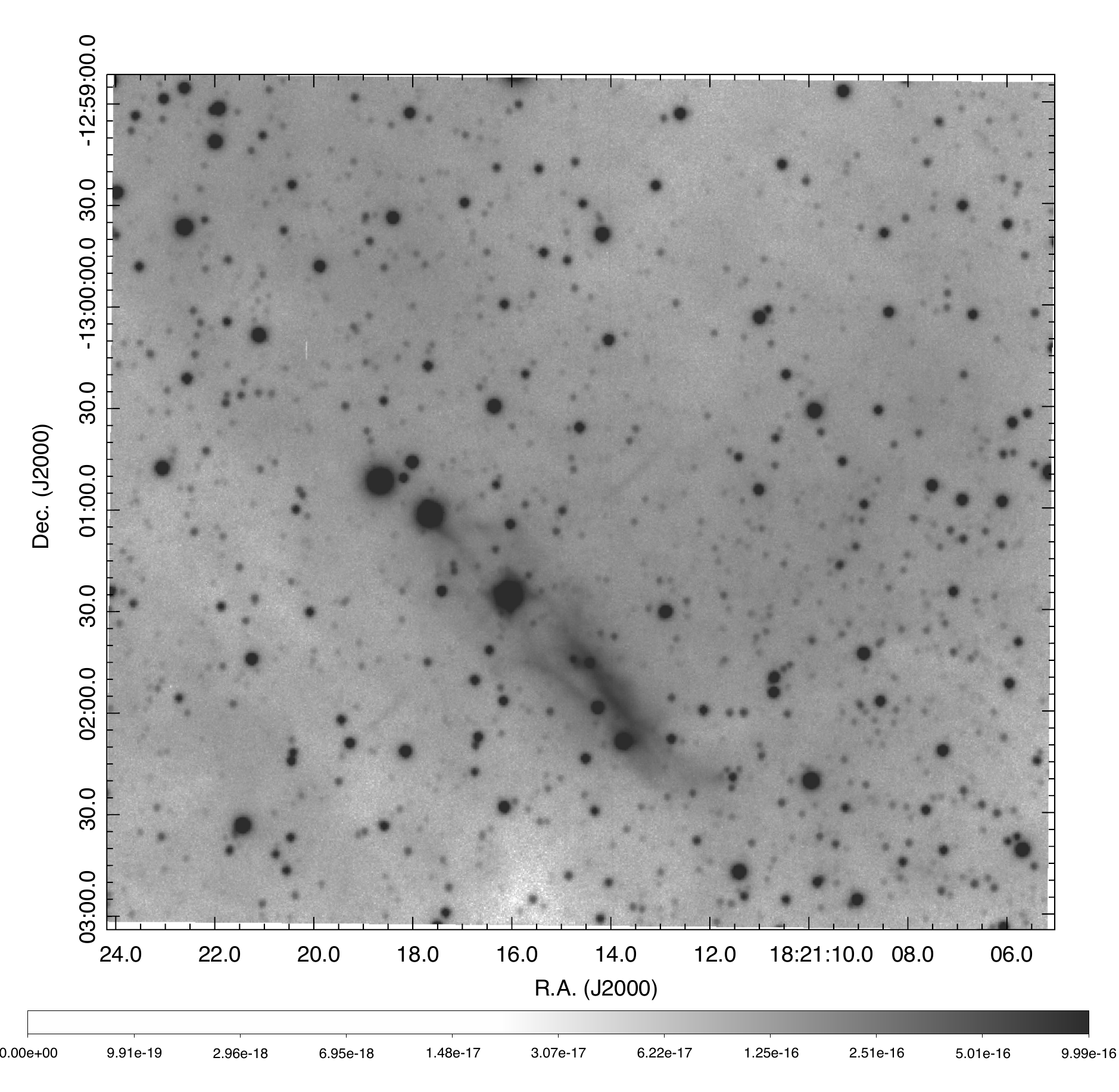}
    \caption{The Red Square Nebula showing  [\Nii ]  emission using a logarithmic display and  shown without labels to highlight the brighter components of the extended MWC922 nebula. The displayed surface brightness ranges from $0$ to $\rm 1 \times 10^{-15}$ (\SB ).}
    \label{fig3}
\end{figure*}

\begin{figure*}
	\includegraphics[width=5in]{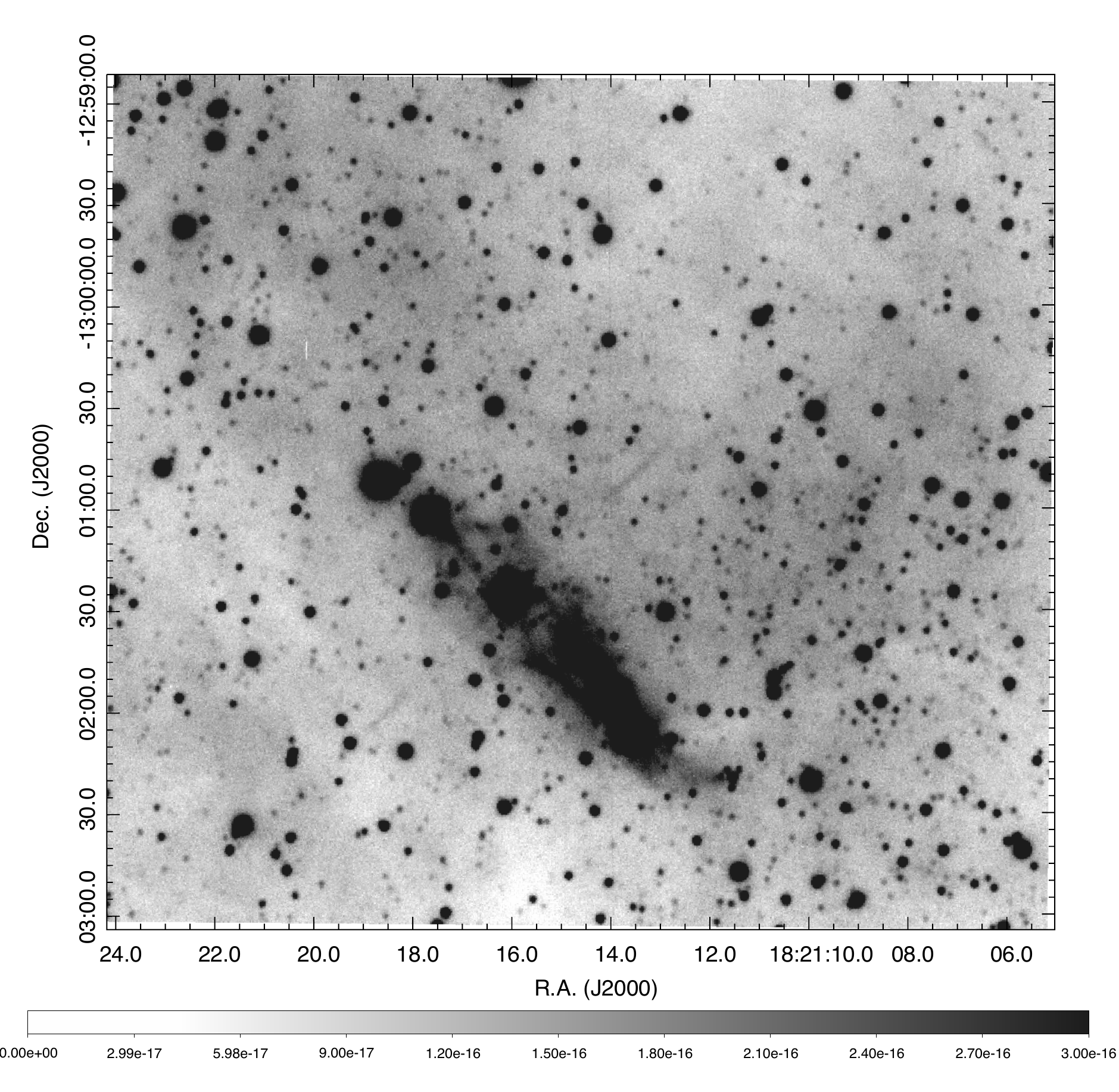}
    \caption{The Red Square Nebula showing the  [\Nii ]  emission using a linear display emphasizing the fainter components of the extended MWC922 nebula.
    The displayed surface brightness ranges from 0 to $\rm 3 \times 10^{-16}$ (\SB ).}
    \label{fig4}
\end{figure*}

\begin{figure*}
	\includegraphics[width=5in]{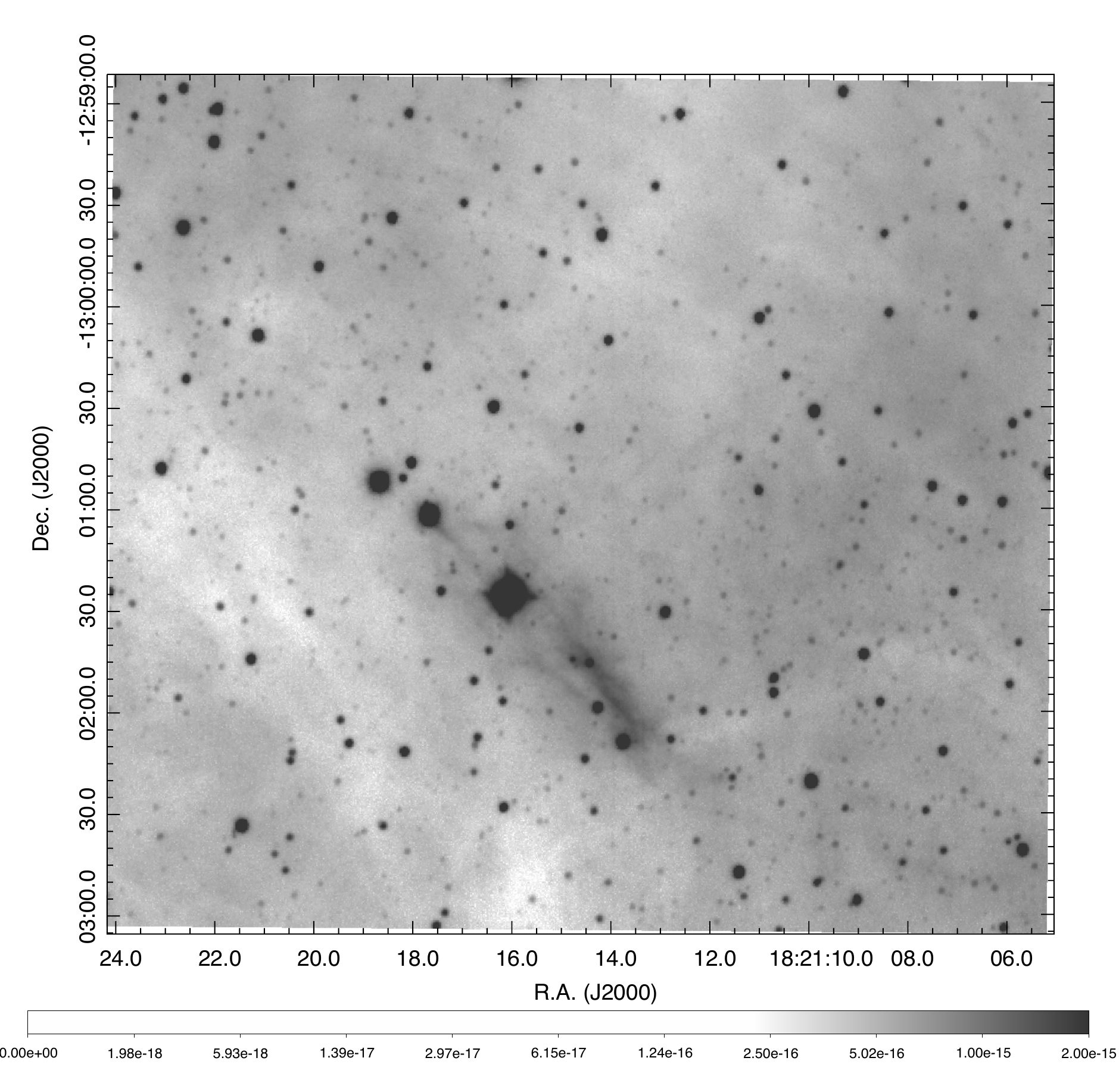}
    \caption{The Red Square Nebula showing the  \Ha\  emission using a logarithmic display showing the full range of nebular emission.
    The displayed surface brightness ranges from 0  to $\rm 2 \times 10^{-15}$ (\SB ).}
    \label{fig5}
\end{figure*}

\begin{figure*}
	\includegraphics[width=5in]{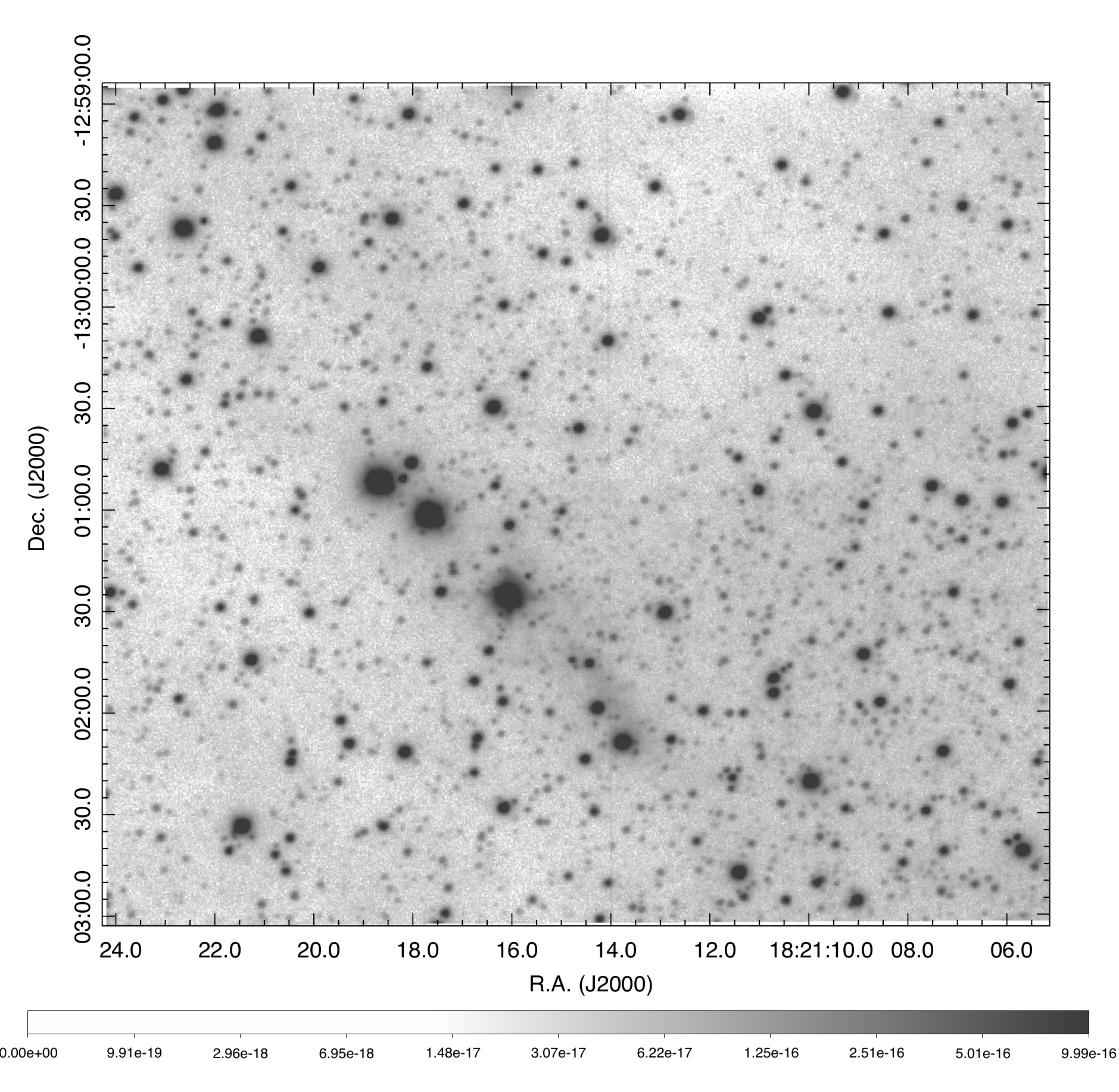}
    \caption{The Red Square Nebula showing the [\Sii ] emission using a logarithmic display.   
    The displayed surface brightness ranges from 0 to $\rm 1 \times 10^{-15}$ (\SB ).}
    \label{fig6}
\end{figure*}

Figures \ref{fig2} and  \ref{fig3} show the outer structure of the Red Square Nebula (RSN) in unprecedented detail in  [\Nii ].   Figure ~\ref{fig2} shows the [\Nii ] image labeled with the various components discussed in the text.  Figure  \ref{fig4} shows the RSN in [\Nii ] in a deep cut, Figure \ref{fig5} shows the RSN in \Ha , and Figure \ref{fig6} shows the same field in [\Sii ].    The brightest part of the nebula consists of an approximately 2\arcmin  -long northeast-southwest oriented `bar' of emission seen in both H$\alpha$ and 6584 \AA\ [NII] emission (`SW disk').  This is the feature detected by \citet{Marston2008}.  The southwest portion facing Messier 16 is much brighter than the northeast portion and is split by a dark band of obscuration that tapers from a width of about 5\arcsec\ near MWC~922 to about 2\arcsec\ at its southwestern edge located about 60\arcsec\ from MWC~922.   The northwest-facing surface of this band is brighter and thicker in both \Ha\ and [\Nii ] than its southeast-facing surface.    These two surfaces merge into a dim, luminous extension which bends due west about 70\arcsec\  of MWC~922 and can be traced for another 19\arcsec\ in this direction as a bent `tail' (`SW tail').  As discussed below, the dark band likely traces a nearly edge-on disk  shed by MWC~922 whose surface layers are externally ionized by the ambient radiation field.  

Towards the northeast, in the direction facing away from Messier 16, the emission is much dimmer (`NE disk').  It is possible that the `SW disk' and `SW tail' discussed above shield the `NE disk'.   Interior to the `SW disk' and `NE disk' but outside the RSN, there is a dim, roughly hexagonal region approximately 40\arcsec\ in diameter (`Hexagon' in Figure \ref{fig2}).   The hexagon is  centered on the Red Square whose bright emission forms a 11\arcsec\ diameter square centered on MWC~922 (shown in Figure \ref{fig1}).  The outer edges of the hexagon towards the north and east are slightly brighter than the interior or the southeast and northwest-facing edges. 

There is a highly collimated, segmented, bipolar  jet emerging from MWC~922 along a southeast-northwest axis at PA $\approx$134\degr\ (towards the southeast) and $\approx$316\degr\ (towards the northwest).   The jet is oriented perpendicular to the edge-on disk whose disk plane has position angle 46\degr\ \citep{Tuthill_Lloyd2007} and the extended nebulae extending towards the southwest in the RSN to within 2\degr .   The innermost segments, NW1 and SE1, emerge from the RSN and can be traced for about 18\arcsec\ and 16\arcsec\ from MWC~922.  These segments become progressively dimmer with distance from the star.     Beyond these inner jet segments, there is a gap extending from 17\arcsec\ to about 40\arcsec\ from the star towards both the southeast and northwest.   The jet becomes visible again from 40\arcsec\ to 56\arcsec\ from MWC~922 towards the southeast (jet segments `SE2').   A corresponding segment extends from 44\arcsec\ to 68\arcsec\ towards the northwest (`NW2').    The most distant parts of the jet are marginally resolved with an observed  width of 1\arcsec\  to 2\arcsec .    Deconvolution of the $\sim$1\arcsec\ seeing disk from the 2\arcsec\  width about 1\arcmin\ from MWC~922 results in an estimate of  the jet opening angle, about  1.5 to 1.7\degr .   These jet segments have similar structure and surface brightness in the \Ha\ and [\Nii ] images and are not seen at our level of sensitivity in [\Sii ].    However, the base of the jet is faintly detected as a filament extending up to 15\arcsec\ from MWC~922 towards both the northwest and southeast with a surface brightness of about $10^{-17}$ \ergsec\  in [\Sii ] (Figure \ref{fig6}). 

\begin{table*}
	\centering
	\caption{MWC922 Dimensions, Surface Brightness, and Derived Parameters}
	\label{table2}
	\begin{tabular}{lcccccccccl} 
		\hline
      Name 		& D$^a$		& PA$^b$   	& R$^c$	& t$^d_{dyn}$	&	SB$^e$ 	& EM(0)$^f$  		& EM(2)$^g$    	& n$_e$$^h$ 		& L$^i$			& Comments \\
      			& (\arcsec  )	& (\degr )	        & (pc)	&   (year)	 	&   ($^e$)   		&  (cm$^6$ pc) 	& (cm$^6$ pc) 	& (cm$^{-3}$) 	& (\arcsec ) 	&  \\
		\hline \hline
SE1 			&  0 - 18		&  134 		& 0 - 0.15 		& 290 	& 0.17 		& 	11 		& 46 			& 36 - 74		& 1 			& from [\Nii ] \\ 

Gap SE1		& 18 - 40		&    - 		& 0.15 - 0.34	& 645	& -	   		& -      		&	-		&	-		&	-		& Gap \\

SE2			& 40 - 56		&  134		& 0.34 - 0.48	& 903	& 0.37 		& 18			& 77	 		& 38 - 78		& 1.5 		& from \Ha  \\
			&			&			&			&		& 0.27 		& 18			& 74			& 37	- 76		& "			& from [\Nii ] \\
			
RSN			& 11 - 11		&	- 		& 0.09 - 0.09	& -		& 	-		& 	-		& -			&	-		& 	-		& Red Square Nebula \\

NW1			& 0 - 16		&  316		& 0 - 0.14		& 258	& 0.62		& 40			& 69			& 69 - 141		& 1			& from [\Nii ]  \\

Gap NW1		& 17 - 44		& 316 		& 0.14 - 0.37 	& 709	& 	-		& 	-		& -			&	-		& 	-		& Gap \\

NW2			& 44 - 68		& 316		& 0.37 - 0.58	&1095	& 0.35		& 17			& 71			& 37 - 75		& 1.5			& from \Ha   \\
			&			&			&			&		& 0.23		& 15			&63			& 34 - 70		& "			& from [\Nii ] \\
			
NW shock 	& 194		& 316		& 1.65		& 3142	& 1.0			& 51			& 216		& 35 - 71		& 5			&  from \Ha \\
			&			&			&			&		& 0.40		& 26			& 109		& 25	 - 51		&			&  from [\Nii ] \\
			
Hexagon    	& 21 - 21		&    - 		& 0.18		& -		& 1.5			& 72			& 306		& 20 - 40         	& 23			&  from \Ha  \\
			&			&			& 0.18		& -		& 1.5			& 72			& 304		& 19 - 40		&			&  from [\Nii ] \\
			
NE "disk"		& 10 - 33		& 44			& 0.28		& -		& 1.4			& 69			& 291		& 16	- 34		& 30			& from \Ha  \\
			&			&			&			& -		& 1.05		& 68			& 287		& 16	-34		&			&  from [\Nii ] \\
			
SW "disk"		& 19 - 71		& 224	 	& 0.60		& -		& 6.3			& 305		& 1287		& 35 - 71		& 30			& from \Ha  \\
			&			&			&			& - 		& 5.1			& 331		& 1392		& 36 - 74		&			& from [\Nii ]  \\
			
SW "tail"		& 71 - 90		& $\sim$260	& 0.60 - 0.79	& -		& 2.0			& 97			& 409		& 25 - 50	 	& 19			& from \Ha  \\
			&			&			&			& -		& 1.11		& 72			& 304		& 21 - 43		&			& from [\Nii ]  \\
			
\hline
\multicolumn{11}{l}{$^a$  D is the projected distance from MWC~922 in arc-seconds.} \\
\multicolumn{11}{l}{$^b$  PA is the position angle of the feature measured from North to East.} \\
\multicolumn{11}{l}{$^c$  R is the projected distance from MWC~922 in parsecs assuming a distance of 1.7 kpc.} \\
\multicolumn{11}{l}{$^d$  t$^d_{dyn}$=R / V$_{jet}$ is the dynamical time assuming a jet velocity, V$_{jet}$ = 500 \kms.} \\
\multicolumn{11}{l}{$^e$  SB is the \Ha\  or [\Nii ] surface brightness given in units of  $10^{-16}$ \SB .}  \\
\multicolumn{11}{l}{$^f$  Emission measure assuming A$_V$ = 0 magnitudes.} \\
\multicolumn{11}{l}{$^g$ Emission measure assuming A$_V$ = 2 magnitudes.} \\
\multicolumn{11}{l}{$^h$  Electron density range for A$_V$ = 0 to 2 magnitudes, assuming a line-of-sight depth L from column $i$.} \\
\multicolumn{11}{l}{$^i$  The assumed line of sight depth of the feature expressed as the width of a feature in arc-seconds on the images .} \\
\multicolumn{11}{l}{Note: For features only seen in [\Nii ], electron density was estimated from the [\Nii ] surface brightness multiplied by 1.33,} \\ 
    \multicolumn{11}{l}{the measured \Ha / [\Nii ] intensity ratio.  See text fro details.} \\ 
\end{tabular}
\end{table*}

The most distant feature potentially associated with the jet from MWC~922 is a faint 1\arcsec\ by 5\arcsec\ emission line feature visible in both \Ha\ and [\Nii ] but not in [\Sii ] dubbed `NW shock'.   `NW shock' is located 194\arcsec\ northwest from MWC~922 and its centroid is at position angle PA = 316\degr\ with respect to the source star.     It lies on the northwest jet axis and its long dimension is at right angles to the jet.   The angular diameter of this feature as seen from MWC~922 is about 3\degr .    The complete absence in our [\Sii ] images  which were obtained with a filter having over three times the bandpass of the \Ha\ and [\Nii ] filters makes it unlikely to be a chance superposition of unresolved, faint stellar images.  Thus, the NW shock is a pure emission line object and, given its location on the northwest jet axis,  may be a part of the MWC~922 outflow.   If this is a bow shock where the jet impacts slower-moving or stationary material,  sideways splashing of plasma entering the shock may explain its angular size as seen from the jet source.  The \Ha\ divided by  [\Nii ] flux ratio is about 2.5, lower than the values in the various parts of the RSN and close to typical \Hii\ region values.   No counterpart is seen towards the southeast axis.   

Table~2 lists some of the observed and derived characteristics of features described in this section and marked in Figure \ref{fig2}.   The contrast between the jet segments,  the terminal bow shock, and the diffuse background which fills the images  is greatest in the [\Nii ] image, making the jet most visible there.  The I(\Ha ) $/$ I([\Nii ] ) flux-ratio is around $1\pm 0.3$ in the jet segments.  In the northeast-southwest  RSN, this ratio is around  $1.35 \pm 0.1$, much higher than in \Hii\ regions.   On the other hand, the  I(\Ha ) $/$ I([\Sii ]) ratio in the RSN is around $8.4 \pm 0.5$, similar to the values  in  \Hii\ regions.     The jet and terminal bow are not seen in the [\Sii ] image, indicating that the flux in \Ha\ is  at least about an order of magnitude larger than in [\Sii ]. 

Table~2 gives two sets of entries for most features with the first being measured on the \Ha\ image and the second estimated from the [\Nii ] image.   Some features such as the inner jet segments SE1 and NW1 are only clearly seen in [\Nii ], likely because the background from the extended \Ha\ emission is so bright.     The sixth column (superscript e) in Table~2 gives the measured surface brightness values for \Ha\ and [\Nii ] for various components of the RSN.  Note that these are {\it differential} measurements in which an OFF-source value is subtracted from the ON-source value using the same aperture.  The [\Nii ] surface brightness values are multiplied by 1.33, the measured mean \Ha / [\Nii ] surface brightness ratio, to estimate the expected \Ha\ surface brightness values.     These values are then used  to compute emission measures and electron densities under two assumptions about foreground extinction; $A_V$ = 0 or $A_V$ = 2 magnitudes.   For the electron density estimation, it is assumed that the line-of-sight depth, $L$, of the emission region is comparable to the jet width for the jet segments, or the maximum length of the emission for  the extended RSN "disk" segments.  The value used for the path length $L$ is given in column i in Table~2.

\section{Discussion}

In this section, the observations are used to deduce the jet velocity and mechanical parameters followed by a discussion of the likely properties of the excretion disk surrounding MWC~922, and the relationship to the Serpens OB1 or OB2 associations and the major star forming complexes Messier 16 and Sh2-54.

\subsection{Jet Physical Properties Determined from \Ha\  Emission}

The \Ha\ surface brightness of the jet segments can be estimated by using the photometric zeropoint (ZPT) of the images.    A rectangular aperture surrounding each region of interest  is used to measure the total counts on the  labeled jet segments  of the MWC~922 system (ON).  The sky-level in the surrounding region is measured using the same aperture (OFF).  The difference in counts, ON - OFF, is used to compute the total magnitude of an extended nebular feature, 
$$
m_{\nu} = -2.5 ~log_{10} [ (ON - OFF) / t_{exp}] + ZPT
$$
where $ t_{exp}$ is the effective exposure time (ZPT here is in units of counts per second for the telescope
and CCD).

The flux in the aperture from each region of interest is then,
$$
F = 1.0 \times 10^{-23}  F_{\nu}(Vega)  \Delta \nu  ~e^{\tau _{\nu}} ~10^{-m_{\nu} / 2.5}  ~~ (erg~s^{-1}~cm^{-2})
$$
where $F_{\nu}(Vega)$ is the flux-density of Vega in the photometric system in which the stellar magnitudes of in-field standards were measured (SDSS r or Johnson R) in Janskys (Jy).   $\Delta \nu$ is the bandpass of the filter (30\AA\ or 100\AA\ in these observations)  expressed in Hz units.  $\tau _{\nu}$ is the optical depth of foreground dust,  $\tau _{\nu} = A_{\nu} / 1.086$, where  $A_{\nu}$ is determined from the interstellar (ISM) reddening curve at $\nu = c / \lambda$, the frequency of the \Ha\ emission line.    Fitting the standard reddening curve for $R_V \approx 3.1$ between $\lambda$ = 5500\AA\  at which the extinction $A_V$ is usually quoted and the 6563\AA\ wavelength of \Ha\ implies a wavelength dependence of $A_{\lambda} / A_V$ proportional to $\lambda ^{-1.4}$ implying that $A_{H \alpha} = 0.78 A_V $. 

The surface brightness of an extended nebular emission feature is then 
$$
SB = F / A     ~~~~~~~~ (erg~s^{-1}~cm^{-2}~arcsec^{-2})
$$  
 where $A$ is the projected area of the measurement box enclosing the radiating feature measured in square arcseconds.

The emission measure  is related to the  surface brightness, $SB$, by
$$
EM = n_e^2  L_{pc} = 4.86 \times 10^{17} SB \times T_4^{0.9} ~~~~  (cm^{-6} pc) 
$$
where $n_e$ is the electron density (in $cm^{-3}$) and $L_{pc}$ is the depth of the emitting region along the line-of-sight (LOS) in units of parsecs (pc).   
Here $T_4$ is the plasma temperature in units of $10^4$ K \citep{Haffner1998}. Under the assumption that the LOS depth is identical to the spatial extent of the emission region on the plane of the sky, this formula can be inverted to solve for the electron density of the radiating plasma, assuming it is fully ionized:
$$
n_e = (EM / L_{pc})^{1/2}   ~~~~~~~(cm^{-3}).
$$

Under these assumptions, the mass per unit area of the emission region $M/A$, and for material moving with a velocity $V$, the mass loss rate $\dot M$ associated with the flow, the momentum injection rate $\dot P$,  and the mechanical luminosity (kinetic energy per second) $\dot E$ can be estimated. 

Given the low-surface brightness of the MWC~922 jet, we have no direct measure of its radial velocity or internal velocity dispersion.  Furthermore, given the detected disk orientation which is nearly edge-on, we expect the predominant motion of the jet to be in the plane of the sky.  However, since this is the first detection of the jet and the jet is too faint to have been detected on previous images, no proper motion measurements are available.  

Therefore, we estimate the jet velocity from the opening angle of the northwest jet as seen from MWC~922.  Photo-ionized hydrogen-dominated plasmas tend to be thermostated to temperatures in the range of 5,000 to 10,000 Kelvin when a near-Solar abundance of heavy elements are present.  Emission lines of relatively abundant trace ions and neutrals such as [\Oi  ], [\Oii ], [\Oiii ], and [\Sii ] which have $\sim$2 eV transitions tend to set the temperature in this range.  The internal sound speed in such a plasma, $c_s = (k T / \mu m_H)^{1/2}$, is around 5.4 to 11.0 \kms\ for a mean molecular weight $\mu = 1.4$ for mostly neutral atomic gas at T = 5,000 K, to $\mu = 0.7$  for fully ionized plasma at $10^4$ K.  

Freely expanding thermal jets unconfined by the pressure of the ambient medium or magnetic fields will expand at the Mach angle given by 
$$
\theta _{Mach} \approx 2 c_s / V_{jet}.
$$
Using the estimated opening angle of the jet discussed above of 1.5 to 1.7\degr , combined with the sound speed estimated from an assumed temperature of 5,000 to 10,000 K implies a jet velocity of
$V_{jet}$ = 360 to 840 \kms .    Given an estimate for the jet velocity, $V_{jet}$, the jet mass loss rate is
$$
\dot M = \mu m_H n_e V_{jet} \pi r^2 _{jet} 
$$
where $r_{jet}$ is the jet radius which, assuming cylindrical symmetry, is given by $L_{pc} / 2$, and $V_{jet}$ is the estimated jet velocity.    The momentum injection rate, and kinetic energy injection rate (mechanical luminosity) are
given by : 
$$
\dot P  = \dot M V_{jet} 
$$
$$
\dot E = (1/2) \dot M V^2_{jet}.
$$
The emission measure, EM, the assumed LOS path-length, and resulting estimates for the electron densities are given in Table~2. 
For an electron density ranging from 50 to 100 cm$^{-3}$, an estimated jet speed of 500 \kms ,  mean jet diameter of 1\arcsec\ (= 1700 AU), and A$_V$ = 2 magnitudes of foreground extinction, the mass loss rate, momentum injection rate, and mechanical luminosity of the jet segments are 
$\dot M \sim 5 - 10 \times 10^{-8}$ \Msol ~yr$^{-1}$, 
$\dot P \sim 2.5 - 5 \times 10^{-5}$ \Msol ~\kms ~yr$^{-1}$,  and 
$\dot E \sim 4 - 8 \times 10^{33}$ erg~s$^{-1}$ = 1 - 2  \Lsol  , respectively.

MWC~922 exhibits the characteristics of both an evolved red giant or supergiant and a hot star.    \citet{Tuthill_Lloyd2007} quote a spectral type of B3 to B6.   The strong IR-excess, however, indicates the  presence of dust typically found in an evolved object.    The detection of a circumstellar disk and a highly collimated jet emerging
at right angles to the disk provides indirect evidence that MWC~922 is a symbiotic binary star in which mass-transfer onto a hot companion is occurring via Roche-lobe overflow.  Such a process can produce both an accretion disk accompanied by jet production as well as an expanding excretion disk.     

The above mass loss rate in the jet is about two magnitudes lower than the rate of $\dot M \sim 2 \times 10^{-6}$ \Msol yr$^{-1} (V_{exp} / 5~ km~s^{-1})$ 
estimated by \citet{Sanchez_Contreras2017} from the ionized component within a few hundred AU of MWC~922.    Sculpting of the inner Red Square Nebula likely requires such a high rate, which may also be responsible for the extended RSN and the `hexagon' structure.  The highly-collimated jet  likely represents the fast, inner core of the much slower wide-angle outflow emerging orthogonal to the orbit plane or the suspected binary that lies in the center of the Red Square.    The wide-angle flow is likely to have a much lower outflow speed and higher mass loss rate than the jet.  \citet{Sanchez_Contreras2017} claim a 5 \kms\ expansion speed for the ionized flow at its $\sim$150 AU outer radius.

For a jet velocity of $V_{jet}$=500 \kms , the time taken to cross a distance, $D$, is the dynamical time given by $t_{dyn} = D/V_{jet}$.   This quantity is tabulated in Table \ref{table2} for each jet segment and gap and the suspected terminal bow shock.   The time scales for the ejecta in the disk-plane are more difficult to ascertain since there is no reliable  measure of the current or past wind velocity.    Using V$\sim$5 \kms\ for the free-expansion speed for a dense wind implies a dynamical time-scale of $\sim$10$^5$ years for the $\sim$0.5 pc extension of the `SW disk',   $\sim$1.5$\times$10$^4$  years for the $\sim$0.09 pc radius  `hexagon', and $\sim$0.8$\times$10$^4$ years for the bright, inner RSN.  

Assuming full ionization as the jet emerges from MWC~922, the recombination time, $t_{rec} = 1 / n_e \alpha_B$,  and recombination distance, $L_{rec} \sim V_{jet} t_{rec} = V_{jet} / n_e \alpha_B$, can be checked against the distance (from MWC~922) of the most distant part of the jet, about $\sim$0.6 pc   ($\alpha _B \approx 2.6 \times 10^{-13}$ {$\rm cm^3 s^{-1}$} is the case-B recombination coefficient for hydrogen at $10^4$ K).  Taking a mean jet velocity of $V_{jet}$ = 500 \kms\ implies an electron density (assuming full ionization near the base of the jet) of $n_e \sim$ 50 cm$^{-3}$ which implies a recombination time-scale of order $t_{rec} \sim $ 2,400  years.  At this jet speed, and in the absence of an external UV radiation field, the beam would recombine at a distance $L_{rec} \sim$ 1.2 pc.   Thus, it is possible that the large gap between the northwest end of jet segment NW2 and the suspected NW shock is primarily due to recombination of the jet beam.  However, the  \Ha\ and [\Nii ] emission from the disk surface implies the presence of an external ionizing radiation field.   

Because of thermal expansion of the jet beam at the Mach angle described above, its density, emission measure, and therefore the \Ha\ surface brightness, are expected to decline with increasing distance from the source.  The density of a steady, constant velocity jet spreading into a cone with a constant opening angle is expected to decline as $d^{-2}$, where $d$ is the distance from the point of origin.  The surface brightness (SB) in recombination lines such as \Ha\ scales as the emission measure, EM.  Since the path length through the emission region, $L$, will be proportional to  $d$ for a constant jet opening angle, the SB is expected to decline as $d^{-3}$.    Thus, jet beam spreading is likely to be a better explanation for the disappearance of the MWC~922 jet beyond the segments NW2 and SE2.  

\subsection{The Extended Red Square Nebula: Properties Determined from \Ha\  Emission}

The \Ha\ surface brightness of the SW disk region ranges up to I(\Ha ) $\approx 6 \times 10^{-16}$~\ergsec\ which implies an emission measure of EM $\approx$ 1300 \EM .   The peak occurs about 44\arcsec\ from MWC~922 ($\sim$ 0.36 pc).   If the \Ha\ emission traces 
the externally photo-ionized surface of an extended excretion disk, the line-of-sight (LOS) path length is likely to be comparable to the radial distance from the star.   For a foreground extinction of A$_V$ = 2 magnitudes, $n_e \approx (EM / L)^{1/2} = 70$ \cmq .     For A$_V$ = 0 magnitudes, $n_e \approx$36 \cmq .

The gross asymmetry of the northeast-southwest oriented RSN which points within a few degrees of the nearby \Hii\ region Messier 16 suggests another possibility.   Perhaps MWC~922 was dynamically ejected from the cluster NGC~6611 inside Messier 16.     The radial velocity of MWC~922 is about 10 \kms\ higher than the mean radial velocity of the molecular gas associated with Messier 16.   No proper motion measurements are available in GAIA Data Release 2 (DR2)  for MWC~922, likely because the core is a spatially extended source at visual wavelengths.   

Assuming that MWC~922 has a $\sim$10 \kms\ proper motion in the plane of the sky, it would take about 3 Myr to traverse the projected distance of 30 pc from the location of NGC~6611 to the present location of  MWC~922.   If this star were moving towards the northeast, the bright southwest nebula could be a tail of gas and dust shed by
the motion of a mass-losing star through the ISM.    In this model, the southwest nebula is likely to be a cylinder with an LOS thickness
comparable to its projected width on the plane of the sky.    Using an LOS dimension ranging from 3\arcsec\ to 10\arcsec\ implies an electron density $n_e \sim$ 120 - 220 \cmq .   In this interpretation, the `Hexagon' surrounding the inner RSN could be a bow shock generated as a slow wind from MWC~922 encounters the ISM.   

Such a motion transverse to the jet would imply some degree of jet bending.  For a proper motion of 10 \kms\ and a jet velocity of 500 \kms,\ the jet might be bent back by ISM interactions by an angle of order 1\degr\ which cannot be ruled out by the images.   The strongest arguments against such a model is that it requires a stronger Lyman continuum ionizing flux to illuminate the RSN than the edge-on disk model, and that there is no direct evidence for a proper motion of MWC~922. 

\subsection{The Extended Red Square Nebula:  Mid-IR emission}

MWC~922 is a bright and compact source in the mid-IR.  In the Herschel Galactic plane survey, Hi-GAL data \citep{Molinari2010,Molinari2016}, a faint tail of 70 \um\ dust continuum emission extends to the southwest for about 40\arcsec\ and coincides with the `SW disk' in our images.   The peak 70 \um\ surface brightness is about 168 MJy~sr$^{-1}$;   the mean surface brightness in a 29\arcsec\ by 12\arcsec\ box is $\sim$92 MJy~sr$^{-1}$ corresponding to a total flux density of 0.75 Jy after subtraction of a background using the same aperture above and below the SW disk.  The aperture only encloses the outer parts of the disk from $\sim$23\arcsec\ to $\sim$52\arcsec\ from MWC~922 because the intense infrared light from the star swamps the inner disk in the Hi-GAL data.    Although it is possible that this feature is an artifact associated with the extremely bright 70 $\mu$m emission from MWC~922 itself, we consider this unlikely for two reasons: First, no other similarly bright compact source in the Hi-GAL data shows such a feature.  Second, the 70 $\mu$m feature coincides in size, shape, and orientation with the brightest \Ha\ and [\Nii ] emission from the SW disk.   The disk can also be faintly seen at 170 $\mu$m against a very confusing and complex background.  Photometry, after background subtraction using the same aperture, gives a total flux density of 0.52 Jy.  At 250 $\mu$m the flux is less than 16 Jy and is dominated by the complex background of the inner Galactic plane.

Fitting the 70 to 170 $\mu$m flux ratio by a single-temperature Planck function implies a dust temperature of 54 K.    Dust heated by illumination from MWC~922 at this distance, or by the ambient radiation field, is expected to be colder than 30 K.    The high temperature could be an indication that the grain population radiating at 70 $\mu$m consists of very small grains (VSGs).  However, it is more likely that the grains are heated by electron impact and UV associated with the ionized surface layers of the disk traced by \Ha\  and [\Nii ] emission.   Such warm dust with temperatures in the range 50 to over 100 K is commonly seen as diffuse 24 $\mu$m emission closely associated with \Hii \ regions.  However, in the Spitzer Space Observatory  24 $\mu$m images, the region of extended \Ha\ emission southwest of MWC~922 is flooded by the large flux from the star.

Dust mass in the measurement box can be estimated from 
$$
M =   {{1.0 \times 10^{-23} X_{dust} S_{\nu} D^2} 
           \over
         {\kappa  _{\nu}   B_{\nu} (T)}}
$$
where $S_{\nu}$ is the background-subtracted flux density in the measurement area in Janskys, $X_{dust}$ is the gas-to-dust ratio by mass (for the computations, we assume $X_{dust}$ = 100), $D$ is the distance in centimeters, $\kappa  _{\nu} $ is the dust opacity at the measurement frequency, and $B_{\nu} (T)$ is the Planck function at dust temperature $T$.   The values of  $\kappa  _{\nu} $ were interpolated from the tables in \citet{OH94} for MRN distribution of 'naked' grains (column 1 in Table 1 in \citet{OH94}) to frequencies corresponding to wavelengths of 70, 170, and 250 $\mu$m, the effective central wavelengths of the filters on-board the Herschel Space Observatory.    The dust opacities used here are  61.25, 14.03, and 8.0 cm$^2$~g$^{-1}$ at 70, 170, and 250 $\mu$m, respectively.  

The total hydrogen mass, (assuming standard ISM values for the gas:dust ratio) in the 348 square arcsecond measurement box which appears to be clear of the bright emission from MWC~922 and which coincides with the southwest extension to the RSN is estimated $\sim 6 \times 10^{-4}$ \Msol\ at 70 $\mu$m, with at least a factor of about 2 uncertainty.    The mass upper limits set by the complex background at 170 and 250 $\mu$m are $< 2 \times 10^{-3}$ and $< 0.17$ \Msol , respectively.    The 70~$\mu$m measurement (and possibly the 170~$\mu$m) is likely to be a severe lower bound to the mass of the extended SW disk in the RSN since the observed dust emission likely only traces a small, unusually warn  sub-population of dust grains.  An alternative  very crude estimate for the `SW disk' feature can be obtained from the \Ha\ emission.  Assuming that this emission traces a $r_{disk}$ = 0.5 pc outer radius disk with a thickness of $h$=0.05 pc, comparable to the thickness of the \Ha\ emission, and that this disk has a uniform density throughout,  $M = \pi r^2 h \mu m_H n \sim$ 0.1 \Msol\ for $n$= 50 \cmq , the lower bound on the electron density derived from \Ha\ emission, presumably located at the irradiated disk surface.   This estimate is also likely to be a severe lower bound.

\subsection{Where Does the Ionizing Radiation Come From?}

If the northeast-southwest extension of the RSN is indeed a half-parsec radius excretion disk shed by MWC~922,  the emission measure
of order EM$\sim$1300 implies a disk-surface electron density of around 100 \cmq .   The absence of an intensity gradient away from MWC~922 indicates that the ionizing radiation does not come from that star; it must come from the general environment.    At a projected distance of $\sim$30 pc, the NGC 6611 cluster in Messier 16 by itself is unlikely to provide the UV radiation.  Even in the absence of intervening dust, the flux fails by about an order of magnitude to explain the observed emission measure.   It is more likely that OB stars in the extended Serpens OB1 and possibly the Serpens OB2 associations, some of which may be closer to MWC~922, provide the needed ionization.  The \Ha\ images show that the entire field of view is filled with diffuse \Ha\ emission that is probably ionized by the same stars.  

The extended \Ha\ emission over the entire ARCTIC  field of view indicates the presence of ionizing UV in the environment.   The long-slit spectra of the red [\Sii ] doublet was used to estimate the mean electron density using the intensity  of the 6717\AA\  line divided by the intensity of the 6731\AA\ line.  The result is that the diffuse background is in the low-density limit of the [\Sii ] line ratio, implying has $n_e < $ 240  \cmq .   

\begin{figure*}
	\includegraphics[width=5in]{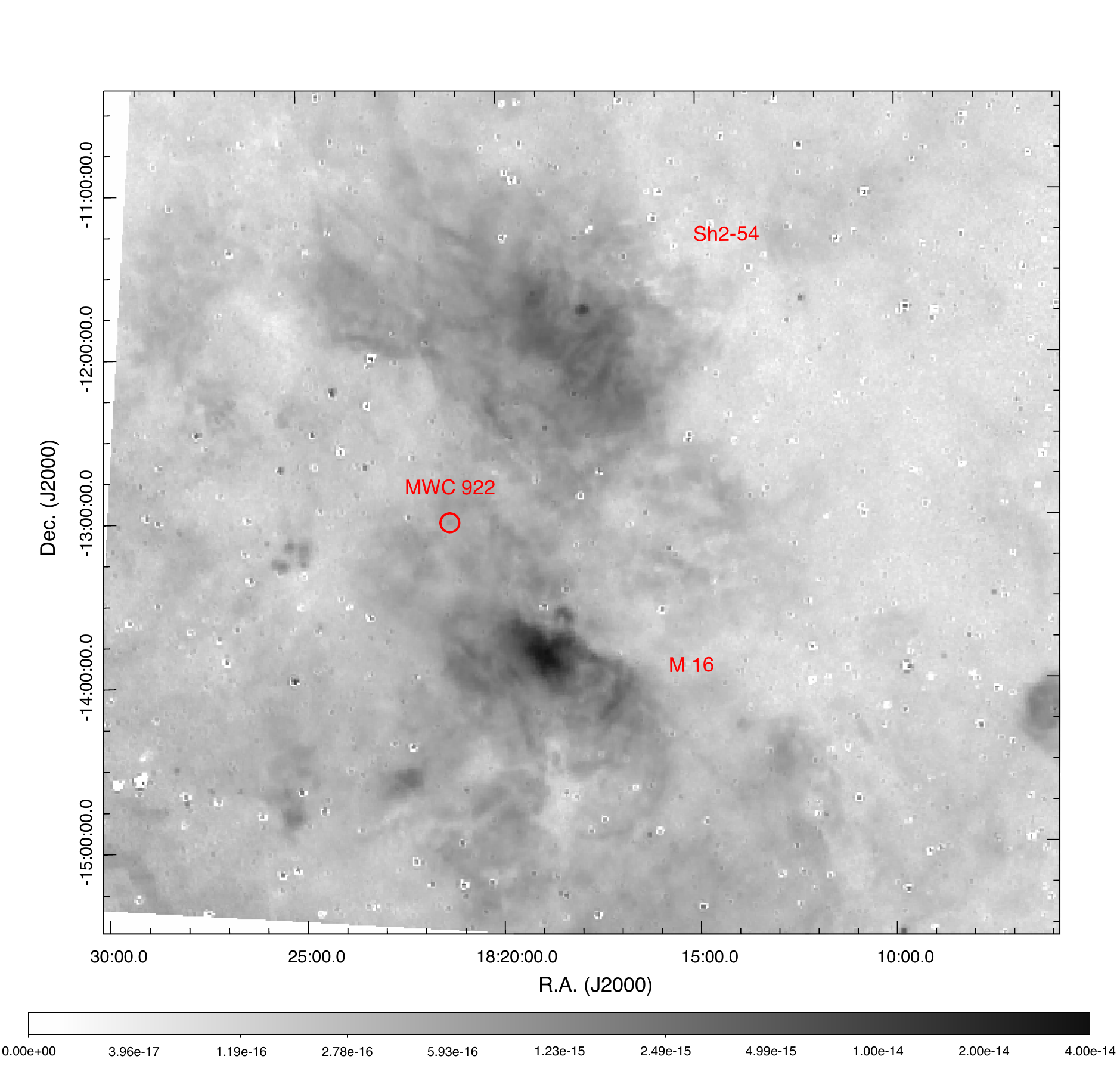}
    \caption{A SHASSA continuum subtracted \Ha\ image showing the environment of the Red Square Nebula (RSN).  The \Hii\  regions Messier 16 and Sh2-54 are indicated along with the RSN.  The surface brightness scales rom of 0 to $\rm 4 \times 10^{-14}$ (\SB ).}
    \label{fig7}
\end{figure*}

Figure \ref{fig7} shows a continuum-subtracted \Ha\ image taken from  SHASSA  \citep{Gaustad2001}.   This image shows a bridge of  emission filling the space between Messier 16 and Sh2-54 with the Red Square Nebula (MWC~922) located along the eastern rim of this feature.    With a mean surface brightness of $5 \times 10^{-16}$\SB\ and a radius of $\sim30$~pc, the mean electron density assuming a uniform-density \Hii\ region and $A_V$ = 0 magnitudes is $n_e \sim$~3 \cmq .  For  $A_V$ = 2 magnitudes, $n_e \sim$~6 \cmq .  The density and radius implies that the Lyman continuum luminosity of ionizing photons is L(LyC)$\sim 1 - 4 \times 10^{49}$ photons~s$^{-1}$ which could be supplied by one or two  late-O-stars, or a dozen early B stars. 

The  low \Ha\ surface brightness, large spatial extent, and  an electron density around $\sim$200 \cmq\ would  indicates that most of the emission likely arrises from a relatively thin layer $L = EM / n^2_e \sim$0.01 to 0.1 pc in depth somewhere along the line of sight.    These layers likely trace ionization fronts at the interface between an extended, extremely low density HII region and surrounding denser clouds.   Ionization is likely powered by a distributed population of massive stars located outside the bright HII regions, Messier 16, Sh2-54, and NGC 6604, and these stars may supply the radiation field illuminating MWC~922.

\section{Conclusions}

Deep, narrow-band images of the Red Square Nebula and its source star, MWC~922, reveal a highly collimated and segmented, parsec-scale  jet oriented orthogonal to the previously identified emission-line nebula which can be traced towards the southwest.    The structure of this externally irradiated jet is used to estimate its speed, found to be around $\sim$500 \kms .   The \Ha\ emission measure is combined with the width of the jet to estimate its electron density which is found to be about $n_e \sim$ 50 to 100 \cmq .  These parameters are used to estimate the mass loss rate, momentum injection rate, and energy injection rate of the jet segments.   These values are $\dot M \sim 5 - 10 \times 10^{-8}$ \Msol ~yr$^{-1}$, 
$\dot P \sim 2.5 - 5 \times 10^{-5}$ \Msol ~\kms ~yr$^{-1}$,  and 
$\dot E \sim 1 - 2$ \Lsol  .

The new images and spectra show that  [\Nii ] emission line at 6584\AA\ is enhanced in the Red Square Nebula.  The jet segments are most visible in [\Nii ], and [\Nii ] is nearly as bright as \Ha\ in the $\sim$0.6 pc-long southwest oriented `tail' of  extended emission which points directly towards Messier~16, located $\sim$30 parsecs to the southwest  in projection.   There is no counterpart to this feature towards the northeast.  However, the images do show a faint, hexagon-shaped nebula surrounding the the bright inner Red Square.  

Two possible models for the southwest facing  nebula  are considered.  It might be a large excretion disk or stream of ejecta shed by MWC~922 which is preferentially illuminated and ionized form the direction of Messier 16.  Because of its orientation, the southwest part shadows the northeast part.   Faint, 70 $\mu$m emission traces warm dust at the surface.   Alternatively, MWC~922 may have been ejected from near Messier~16.  In this model,  the southwest component of the RSN is a tail of ejecta left behind the star as mass lost from the star interacts with the interstellar medium through which it moves.   If the southwest feature is an externally irradiated disk, the \Ha\ emission provides a lower-bound on its mass of order $\sim$0.1 \Msol .

The Red Square Nebula and its jet appears to be externally ionized.   The surrounding interstellar medium is illuminated by diffuse background Lyman continuum.     The extended background emission and ionization of the RSN required a Lyman continuum photon luminosity of order L(LyC) $\sim 1 - 4 \times 10^{49}$ H-ionizing photons per second at a distance of about 30 pc from the RSN.   Such a radiation field could be produced by a distributed population of about a dozen early B stars, or one or two late O stars at about 30 pc distance from the RSN.   

\section*{Acknowledgements}

The work presented here is based on observations obtained with the Apache Point Observatory 3.5-meter telescope, which is owned and operated by the Astrophysical Research Consortium.    We thank the Apache Point Observatory Observing Specialists for their assistance during the observations.   We acknowledge use of 
the Southern H-Alpha Sky Survey Atlas (SHASSA), which is supported by the National Science Foundation \citep{Gaustad2001}.
We thank John (Jack) Faulhaber for useful discussions and comments on the manuscript.  We thank the referee for constructive comments that improved the manuscript.





\bsp	
\label{lastpage}
\end{document}